\documentclass[a4paper,twoside]{article}
\newcommand{\affil}[1]{$^{\rm #1}$}
\usepackage[authoryear]{natbib}
\bibpunct{(}{)}{;}{a}{}{,}
\usepackage{graphicx}
\date{} 
\title{\large\bf\flushleft The Magellanic Clouds as a template for the
study of stellar populations and galaxy interactions}
\author{\parbox{\textwidth}{\flushleft
\vspace{-0.5cm}
{\it M.-R.L. Cioni\affil{A,T}, K. Bekki\affil{B},
  G. Clementini\affil{C}, W.J.G. de Blok\affil{D},
  J.P. Emerson\affil{E}, C.J. Evans\affil{F}, R. de Grijs\affil{G},
  B.K. Gibson\affil{H}, L. Girardi\affil{I}, M.A.T. Groenewegen\affil{J},
  V.D. Ivanov\affil{K}, P. Leisy\affil{L}, M. Marconi\affil{M},
  C. Mastropietro\affil{N}, B. Moore\affil{O}, T. Naylor\affil{P},
  J.M. Oliveira\affil{Q}, V. Ripepi\affil{M}, J.Th. van Loon\affil{Q},
  M.I. Wilkinson\affil{R}, P.R. Wood\affil{S}}\\
\vspace{0.4cm}
{\small \affil{A}\,CAR, University of Hertfordshire, College Lane,
  Hatfield, AL10 9AB, United Kingdom}\\ {\small \affil{B}\,School of
  Physics, Univerisity of New South Wales, Sydney, NSW 2052,
  Australia}\\ {\small \affil{C}\,INAF, Osservatorio Astronomico di
  Bologna, Via Ranzani 1, 40127 Bologna, Italy}\\ {\small
  \affil{D}\,University of Cape Town, Private Bag X3, Rondebosch 7701,
  South Africa}\\ {\small \affil{E}\,Queen Mary University of London,
  Mile End Road, London E1 4NS, United Kingdom}\\ {\small
  \affil{F}\,UK Astronomy Technology Centre, Blackford Hill,
  Edinburgh, EH9 3HJ, United Kingdom}\\ {\small \affil{G}\,Department
  of Physics \& Astronomy, University of Sheffield, Sheffield, S3 7RH,
  United Kingdom}\\ {\small \affil{H}\,Centre for Astrophysics,
  University of Central Lancshire, Preston, PR1 2HE, United Kingdom}\\
  {\small \affil{I}\,INAF, Osservatorio Astronomico di Padova, Vicolo
  dell'Osservatorio 5, 35122 Padova, Italy}\\ {\small
  \affil{J}\,Institute for Astronomy, University of Leuven,
  Celestijnenlaan 200D, 3001 Leuven, Belgium}\\ {\small
  \affil{K}\,European Southern Observatory, Santiago, Av. Alonso de
  C\'{o}rdoba 3107, Casilla 19, Santiago, Chile}\\ {\small
  \affil{L}\,Instituto de Astrof\'{i}sica de Canarias, v\'{i}a
  L\'{a}ctea s/n 38200 La Laguna, Tenerife, Spain}\\ {\small
  \affil{M}\,INAF, Osservatorio Astronomico di Capodimonte, via
  Moiariello 16, 80131 Napoli, Italy}\\ {\small \affil{N}\,University
  Observatory Munich, Wendelstain Observatory, Scheinerstr. 1, 81679
  Munich, Germany}\\ {\small \affil{O}\,Institute for Theoretical
  Physics, University of Zurich, 8057 Zurich, Switzerland}\\ {\small
  \affil{P}\,School of Physics, University of Exeter, Stocker Road,
  Exeter, EX4 4QL, United Kingdom}\\ {\small \affil{Q}\,School of
  Physical and Georgraphical Sciences, University of Keele,
  Staffordshire, ST5 5BG, UK}\\ {\small \affil{R}\,University of
  Leicester, University Road, Leicester, LE1 7RH, United Kingdom}\\
  {\small \affil{S}\,Mount Stromlo Observatory, RSAA, Cotter Road,
  Weston Creek, ACT 2611, Australia}\\ {\small
  \affil{T}\,M.Cioni@herts.ac.uk}}}
\begin{document}
\twocolumn[
\begin{changemargin}{.8cm}{.5cm}
\begin{minipage}{.9\textwidth}
\vspace{-1cm}
\maketitle
%
%
\small{\bf The Magellanic System represents one of the best places to
study the formation and evolution of galaxies.  Photometric surveys of
various depths, areas and wavelengths have had a significant impact on
our understanding of the system; however, a complete picture is still
lacking. VMC (the VISTA near-infrared $YJK_s$ survey of the Magellanic
System) will provide new data to derive the spatially resolved star
formation history and to construct a three-dimensional map of the
system.  These data combined with those from other ongoing and planned
surveys will give us an absolutely unique view of the system opening
up the doors to truly new science!}


\medskip
\medskip
\end{minipage}
\end{changemargin}
]
\small

\section{Introduction}

The Magellanic Clouds (MCs) are interacting SBm galaxies similar to
many that exist in Universe. They are the largest neighbouring
satellites of the Milky Way, reflecting a typical environment of a
large galaxy surrounded by satellites. They contain stars which are as
old as the Universe as well as newly forming and this extended range
of star formation is a highly valuable source to understand the
process of formation and evolution of galaxies in general.

The MCs are overall more metal poor than the Galaxy and therefore may
hold information about the Universe at its early stages. They are
located at a fairly well known distance, which makes it easier to
measure details of their stellar component and structure. They are
also fortunately located in a region of sky only lightly affected by
Galactic reddening, which translates into the capability of detecting
their faint stellar populations.

The MCs belong to a complex system, the Magellanic System, which has
in total four distinct components: the Large Magellanic Cloud
(LMC), the Small Magellanic Cloud (SMC), the Bridge connecting the two
Clouds and the Stream attached to the SMC. The latter two are
predominantly formed of gas and are of tidal origin.

\section{A near-infrared view of the Magellanic Clouds}

The evolved stellar population of the Magellanic Clouds is best
studied in the near-infrared window. At these wavelengths luminous
giant stars: a\-symp\-to\-tic giant branch (AGB) stars and upper red
giant branch (RGB) stars have been detected in large numbers across
the galaxies by wide-field surveys like DENIS and 2MASS.

These surveys have shown that the number density distribution of these
stars traces the morphology and structure of the galaxies. The ratio
between C-rich (or C-type) and O-rich (or M-type) AGB stars, (the C/M
ratio) is an indicator of the iron abundance ([Fe/H]); the relation
between these two quantities has been calibrated using homogeneous
observations of AGB stars in various galaxies of the Local Group
\citep{bd}. The distinction between the two AGB types depends on the
stellar surface chemistry that can be dominated either by carbonaceous
or silicate molecules. These molecules are responsible for the opacity
directly affecting broad-band colours; C-rich AGB stars populate a
large range of $J-K_s$ colours a narrow range of $K_s$ magnitudes
contrary to O-rich AGB stars.  The $K_s$ magnitude distribution of AGB
stars, interpreted using appropriate theoretical models, is a simple
indicator of the mean age and metallicity of the underlying stellar
population showing considerable inhomogeneities across both
galaxies. These results are described in more detail below.

\subsection{The structure of the LMC}

\begin{figure}[h]
\begin{center}
\includegraphics[scale=0.4, angle=0]{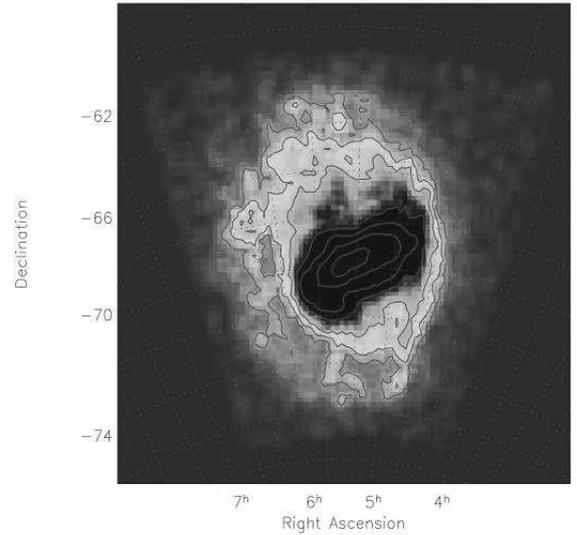}
\caption{Distribution of AGB star counts across the LMC. Contours are
  at $3$, $5$, $10$, $20$, $30$, $50$, $100$, $150$ per $0.04$ deg$^2$
  \citep{mor}.} 
\label{agb}
\end{center}
\end{figure}

The distribution of AGB stars (Fig.~\ref{agb}), regardless of their
spectral type, shows a smooth outer elliptical structure embedding a
thick bar and protuberances emerging from it, hinting at the existence
of spiral arms \citep{mor}.

By selecting AGB stars in a narrow range of colours, their mean
luminosity will trace distances across the structure of the galaxy. In
fact, this method allowed us to derive the orientation of the LMC in
the sky, providing accurate measurements of the inclination and the
position angle \citep{vdm}.

\subsection{The C/M across the LMC}

\begin{figure}[h]
\begin{center}
\includegraphics[scale=0.4, angle=0]{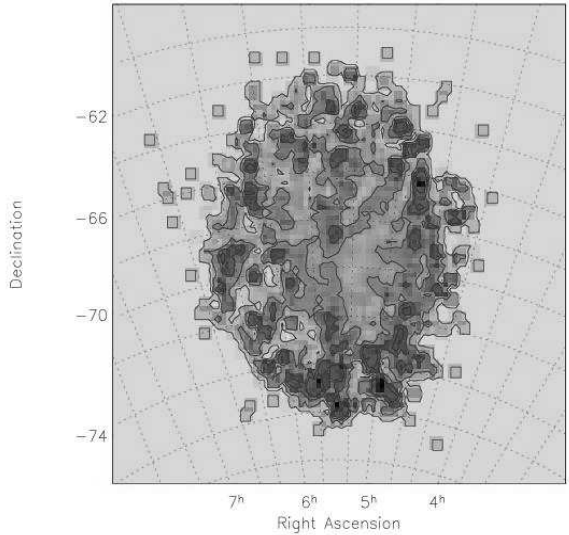}
\caption{C/M ratio distribution across the LMC. Contours are at:
  $0.1$, $0.25$, $0.4$, $0.55$ per $0.04$ deg$^2$ \citep{cm}.}
\label{cm}
\end{center}
\end{figure}

The distribution of the C/M ratio across the LMC (Fig.~\ref{cm}) shows
clearly the existence of the classical metallicity gradient which is
present in many galaxies: the iron abundance is higher in the centre
and decreases more or less radially outwards \citep{cm}.  This trend
has subsequently been confirmed using RGB stars \citep{alves}.

\subsection{The $K_s$ method}

\begin{figure}[h]
\begin{center}
\includegraphics[scale=0.38, angle=0]{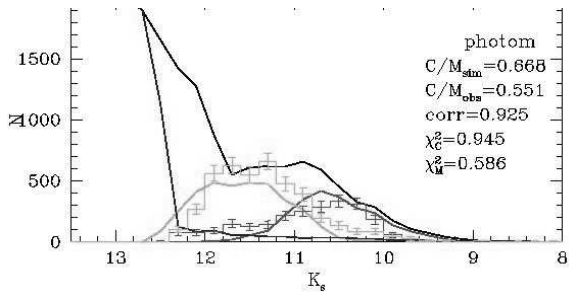}
\caption{Observed magnitude distribution of C and M stars (histograms)
  compared with theoretical distributions (continuous lines) in a
  given region of the LMC \citep{lmc}.}
\label{lumf}
\end{center}
\end{figure}

The $K_s$ magnitude distribution of AGB stars holds important
information on the metallicity and age of the overall stellar
population. In particular, the observed magnitude distribution of
C-rich and O-rich AGB stars can be interpreted using theoretical
distributions constructed using stellar evolutionary models spanning a
range of metallicity and age parameters.  First, it is necessary to
isolate the AGB component from other stellar components such as
foreground stars or stars at other stages of evolution. This is done
using the near-infrared colour-magnitude diagram ($J-K_s$, $K_s$). The
same selection criteria are applied to synthetic diagrams obtained
from stellar evolutionary models. Then, the resulting observed and
theoretical samples are compared to identify the one that best
represents the stellar population in the selected area
(Fig.~\ref{lumf}).

The study of the MCs using this technique with theoretical models from
the Padova group \citep{g00,b94,m99} is presented in
\citet{lmc,smc}. The authors showed that the star formation rate
derived from localised regions within the galaxies does not apply to
the galaxies as a whole but inhomogeneities in both metallicity and
mean age are clearly present. This result is free from systematic
differences that may occur because of the specific stellar models
adopted.

\subsubsection{The LMC mean age and metallicity}
The stellar population across the LMC appears younger in the East than
in the West. The bar has a composite stellar population and does not
show up in the maps of mean age and metallicity. These maps have been
corrected for the orientation of the LMC in the sky and for
foreground, but not differential, reddening.

The C/M ratio traces well the distribution of metallicity obtained
from the $K_s$ method confirming the validity of the C/M ratio as an
indicator of [Fe/H]. There is a region North-East of the centre where
the metallicity is high, as well as in an outer ring-like structure
extending from the South to the West. While the first region probably
corresponds to a place with active star formation, the latter might be
affected by a low number statistics or by the size of the bins adopted
to construct $K_s$ histograms. In fact, the distribution of the C/M
ratio derived in smaller bins (Fig.~\ref{cm}) shows clearly a metal
poor ring around the LMC which surrounds a region where the metal
content is higher. This substructure is washed out by larger bins
necessary to obtain a statistically significant $K_s$ magnitude
distribution for both C-rich and O-rich AGB stars \citep{lmc}.

\subsubsection{The SMC mean age and metallicity}
\begin{figure}[h]
\begin{center}
\includegraphics[scale=0.25, angle=0]{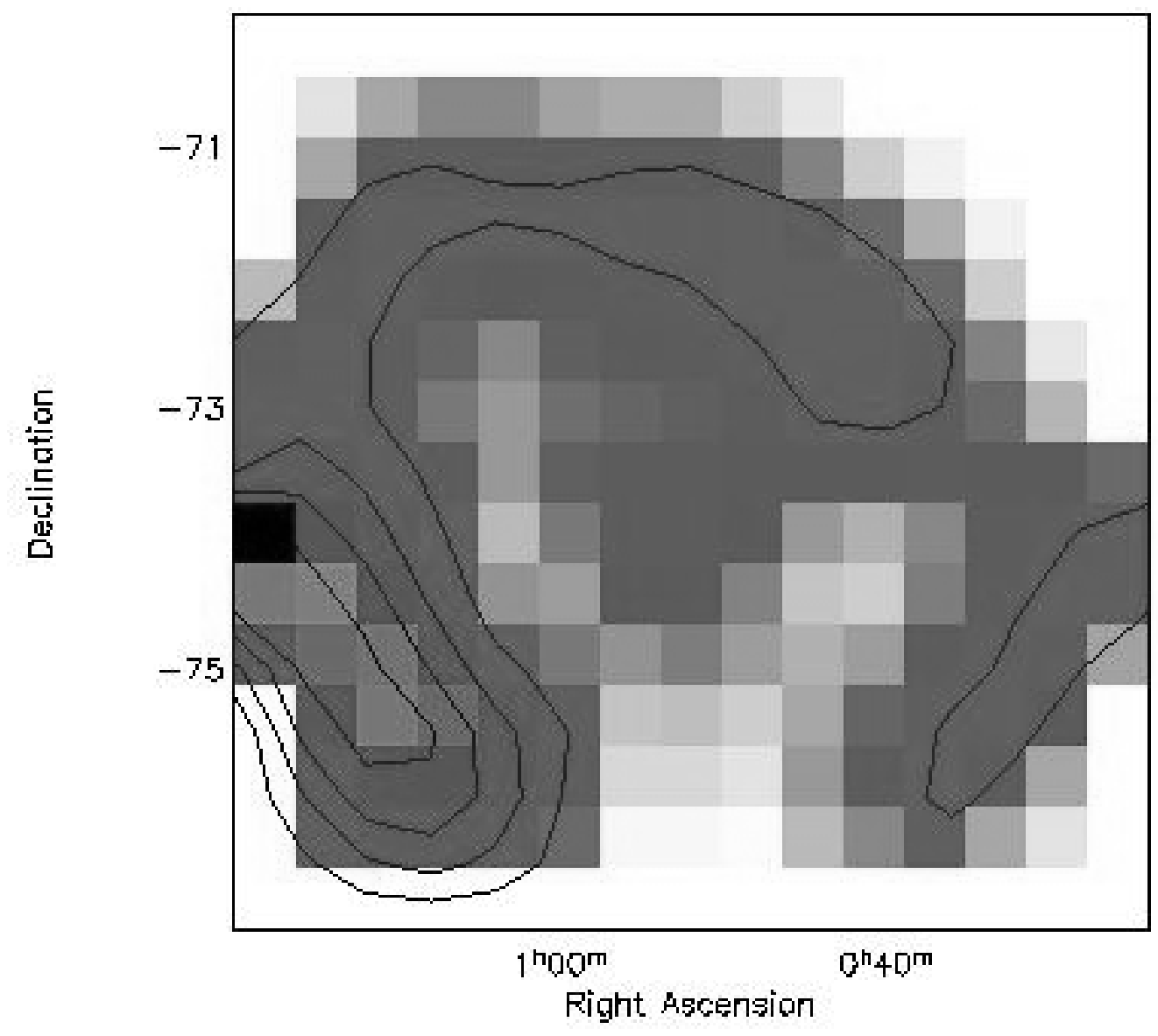}
\includegraphics[scale=0.25, angle=0]{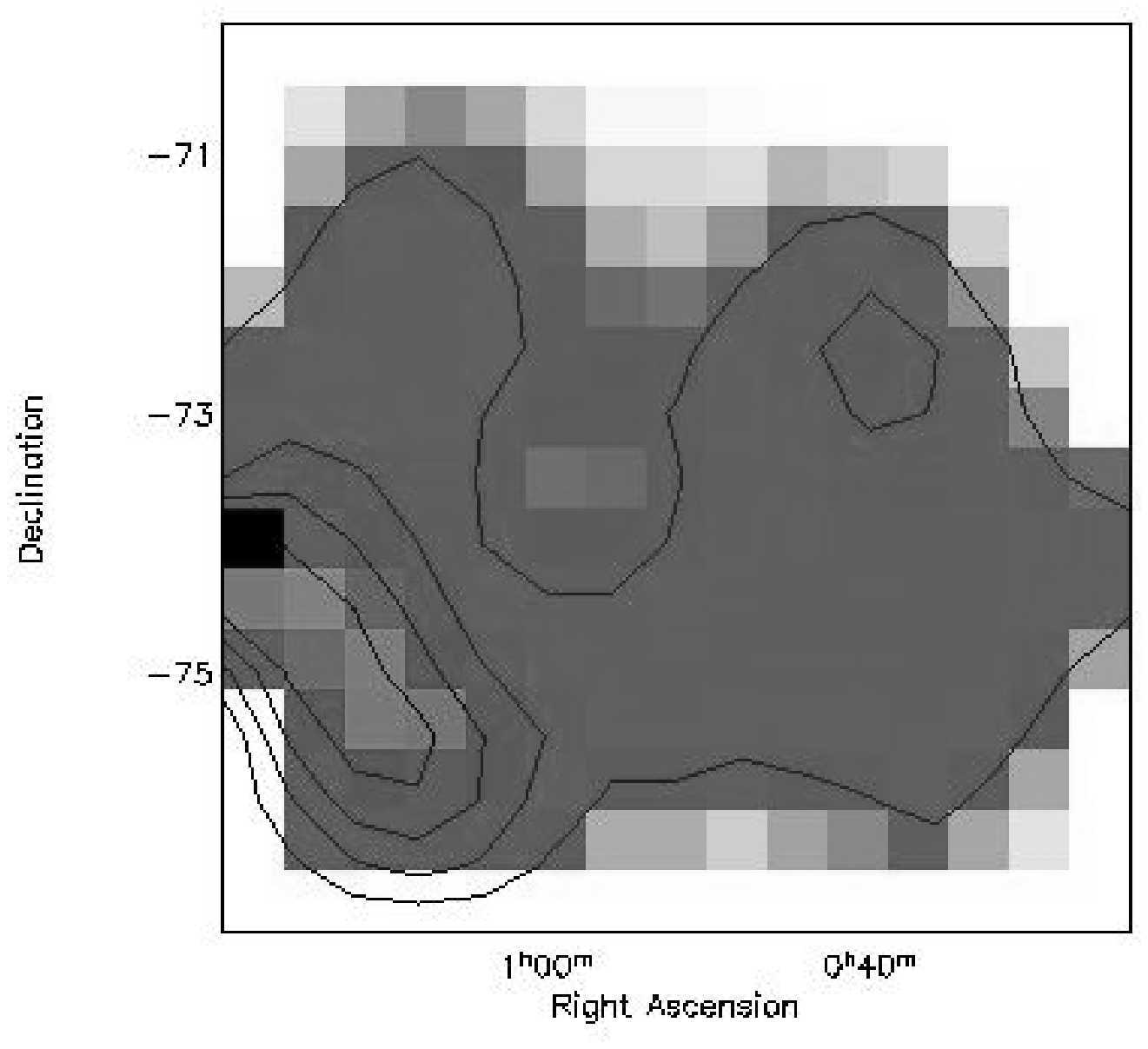}

\includegraphics[scale=0.25, angle=0]{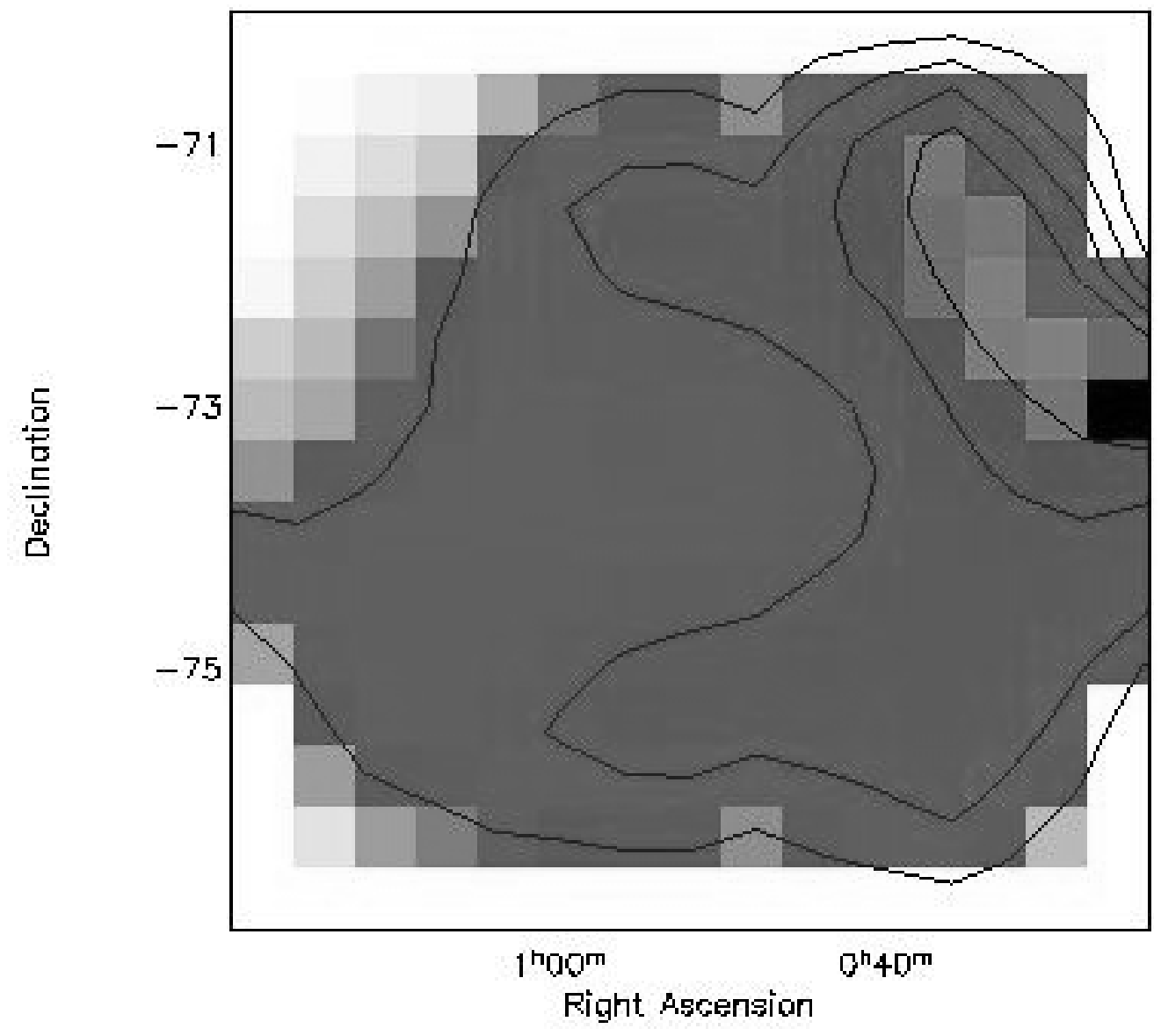}
\includegraphics[scale=0.25, angle=0]{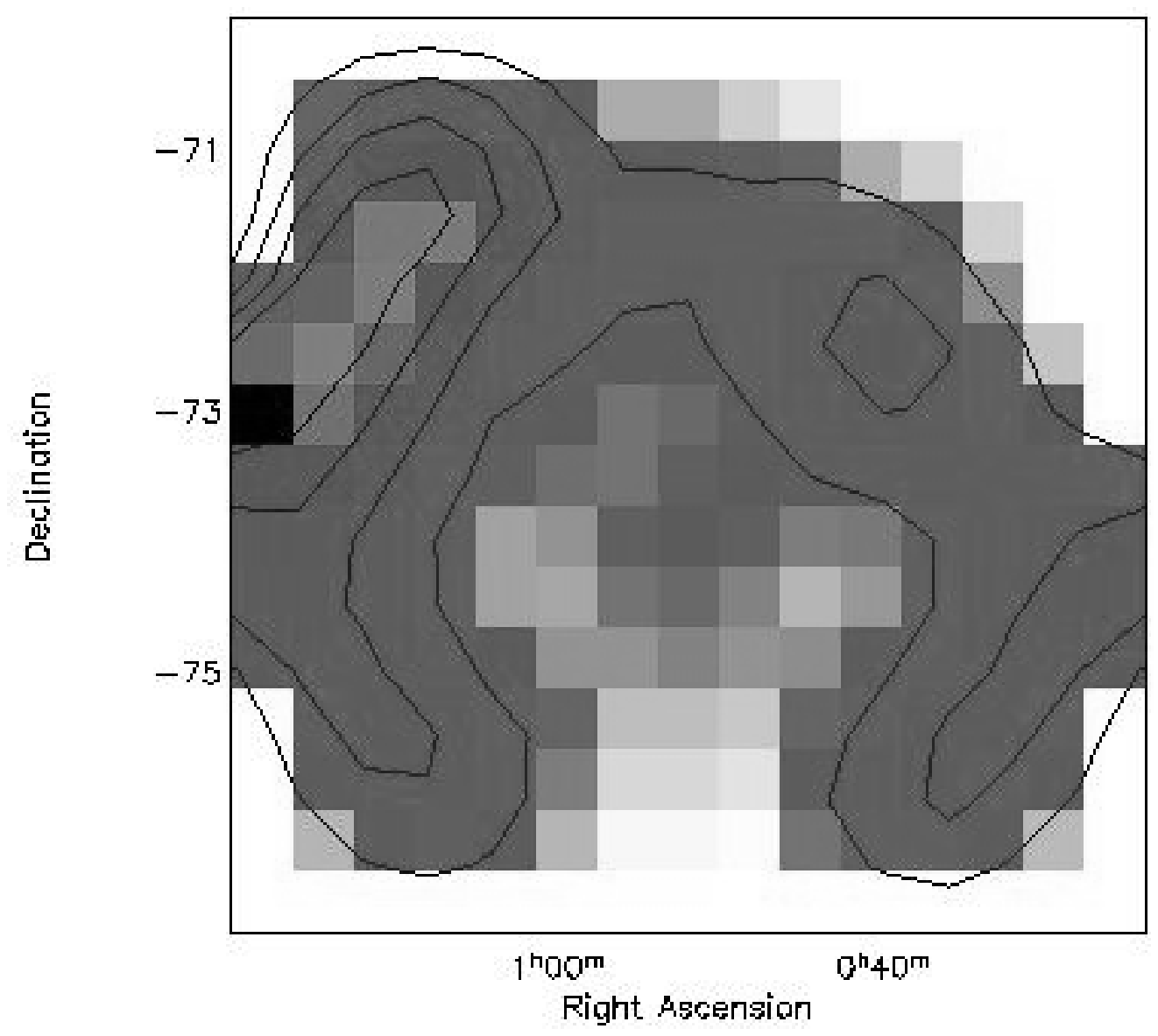}

\includegraphics[scale=0.25, angle=0]{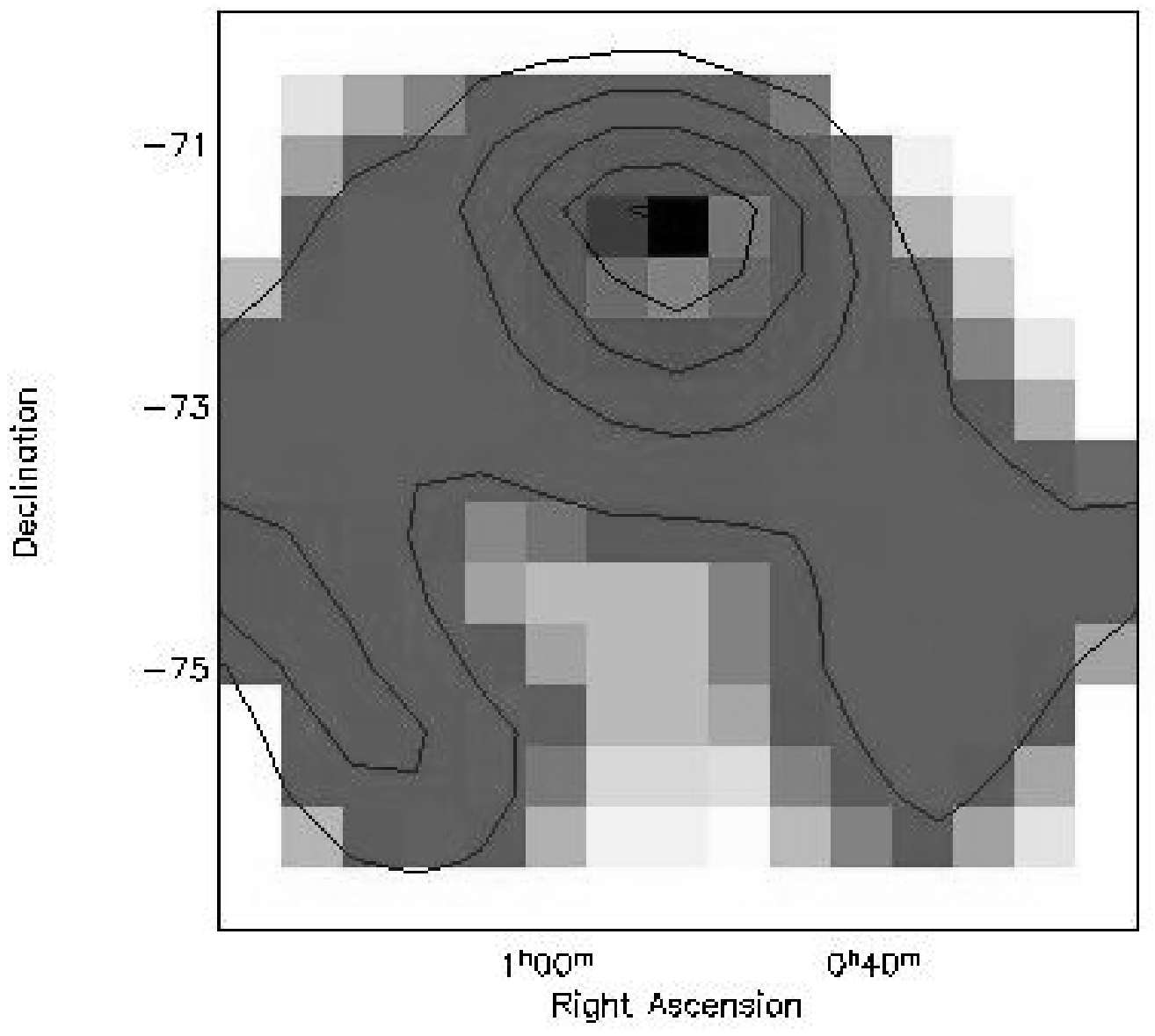}
\caption{Clockwise from the top left: distribution of the metallicity
  that corresponds to the maximum probability for a star
  formation rate equivalent to a mean age of: $2.0$, $3.9$, $6.3$,
  $8.7$ and $10.6$ Gyr for C stars across the SMC. Contours are at: $0.7$,
  $0.8$, $0.9$, $0.94$ and $0.98$ expressed in terms of Z \citep{smc}.}
\label{smc}
\end{center}
\end{figure}

The distribution of metallicity across the SMC as a function of age
shows a very interesting pattern, perhaps associated with the
propagation of star formation throughout time. A region of high
metallicity is located South-East of the galaxy centre and corresponds
to a mean age of $2$ Gyr. Going back in time this region moves
clockwise to the West, leaving the centre of the galaxy metal poor and
with very little variation \citep{smc}, Fig.~\ref{smc}.

A similar ring-like feature traced by high metals was also detected by
\citet{har} in their analysis of the MCs Photometric Survey
da\-ta\-ba\-se. They found that this substructure corresponds to a
stellar population of about $2.5$ Gyr old either originating from a
gas rich merger or awaiting the subsequent inward propagation of star
formation.

\subsection{Origin of inhomogeneities}
Inhomogeneities of metallicity and age are, according to \citet{sim},
fossil records of clumpy pasts of galaxies. This means that stars
form in clumps and each clump has an age and a metallicity. If clumps
are smaller than $10^7$ M$_{\odot}$ then they dissolve to form the
field stellar population. On the contrary, if clumps are larger then
they keep their identity and will form stable structures like galaxy
cores and bars. From what the $K_s$ method has shown us, it is
possible to isolate regions in a galaxy that can be associated with a
dominant mean age and metallicity suggesting that most of the
population there originates from a given clump or from a
combination of clumps which differs from the combination that was
present in another spatially different region.

\begin{figure}[h]
\begin{center}
\includegraphics[scale=0.6, angle=0]{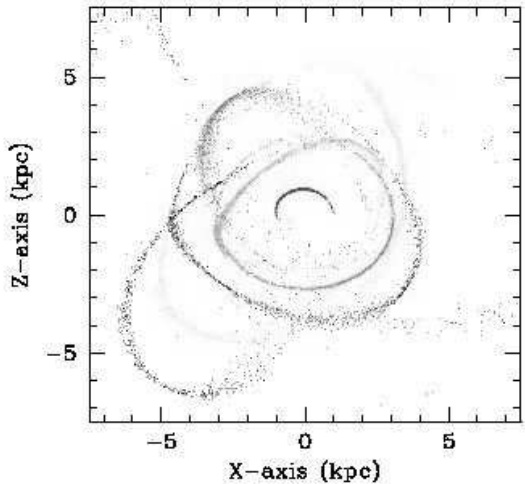}
\caption{Distributions of stars originating from different stellar
  clumps projected onto the $x-z$ plane for different tidal
  interaction models tracing each a different arc; for a colour figure
  see \citep{sim}.}
\label{bekki}
\end{center}
\end{figure}

\subsection{What is missing?}
A complete picture of a galaxy is given by the age, metallicity and
 motion of all constituents at a given place. To obtain these we need
 to observe all objects (stars, dust and gas) and we need
 sophisticated theoretical models that convert observables into
 physical properties (e.g. age, metallicity).

The $K_s$ method, however, provides only average values of the age and
 metallicity. The latter is also measured somewhat more accurately by
 the C/M ratio. What is needed are absolute values and to obtain those
 we need either new observations or, if applicable, more sophisticated 
 analysis techniques and a large effort to develop new theoretical
 models that include and are able to reproduce all observational
 details.

The kinematics is certainly a fundamental aspect which will not be
discussed in more detail in this paper. A knowledge of the full
chemistry is also fundamental, this is not limited to the iron
abundance discussed here but also to the abundance of other elements
such as $\alpha$ elements which play a major role in the evolution
of stars and galaxies.

A knowledge of the structural parameters of galaxies plays an
important role in understanding the distribution of its stellar
content. It is not always easy to disentangle the effect due to
distance and that due to stellar populations from an observed
magnitude shift, especially in galaxies of a moderate size.

\section{The VISTA Public Survey of the Magellanic System}
 
The VISTA public survey of the Magellanic System (VMC), lead by Cioni,
is the result of an international collaboration of astronomers working
in $9$ different countries, but most of them are in the United Kingdom
and Italy. We have prepared a Public Survey proposal to use the newly
developed VISTA telescope with its infrared camera \citep{eme} to
obtain unique observations of the Magellanic System that will
considerably improve our understanding of its formation and evolution
and could be used as a template for the study of galaxy evolution in
general.

\begin{figure*}[t]
\begin{center}
\includegraphics[scale=0.87, angle=0]{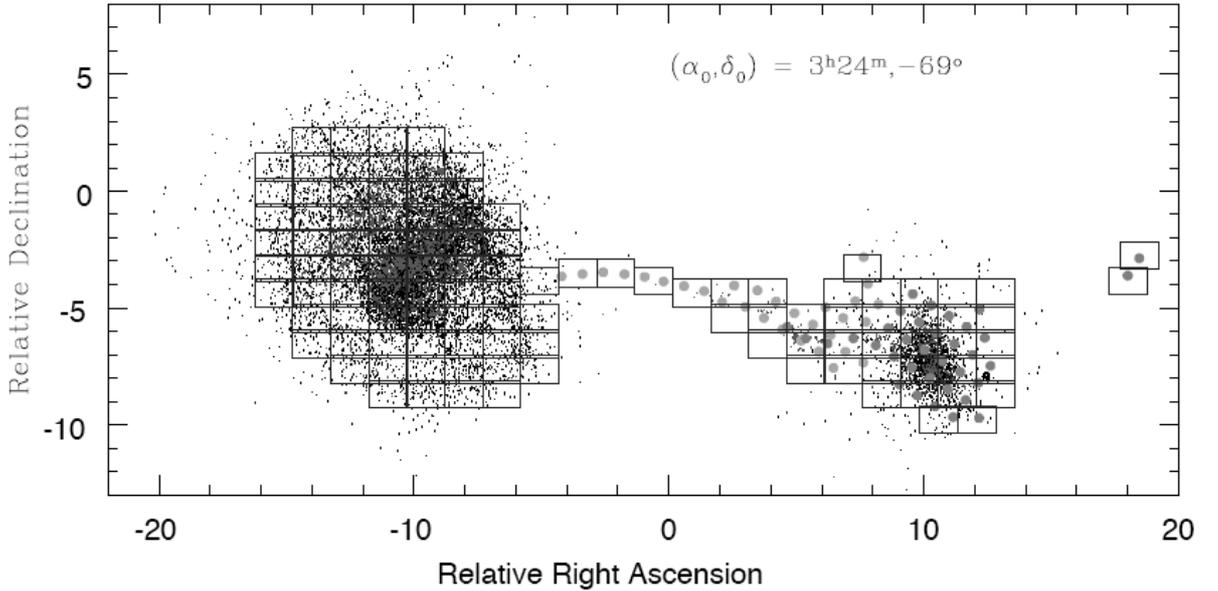}
\caption{Distribution of VISTA tiles across the Magellanic
  System. Underlying small dots indicate the distribution of C stars,
  clusters and associations while thick dots show the location of
  observations to be performed with the VLT Survey Telescope in the
  optical domain.}
\label{vmc}
\end{center}
\end{figure*}

VMC is one of $6$ public surveys approved by the European Southern
Observatory; one of two targeting resolved stellar populations. For
details about VISTA public surveys check the recent publication by
\citet{mes}.

\subsection{VISTA}
The VISTA telescope is the premier infrared survey instrument for the
foreseeable future.  The $4$-m telescope is located at the excellent
Chilean site of Paranal. The $16$ $2048\times2048$ pixel detectors in
the camera cover from $0.82$ to $2.3$ micron in wavelength via (seven)
broad and narrow band filter sets. The pixel size is
$0.339^{\prime\prime}$ with an instrument point spread function of
$0.51^{\prime\prime}$.

In order to cover homogeneously an area of sky observations will be
offset to fill in the gaps among the detectors. These amount to $95$\%
in the $x$ direction and $47.5$\% in the $y$ direction. In particular,
a minimum number of six offsets will cover $1.65$ deg$^2$ where each
pixel will be observed at least twice.

More details about telescope and camera can be found at
http://www.vista.ac.uk/.

\subsection{VMC observing strategy}

VISTA observations of the Magellanic System will be obtained in three
filters: $Y$, $J$ and $K_s$. A total area of $184$ deg$^2$ will be
covered during five years; the survey is expected to begin in the fall
of $2008$. In detail, $116$ deg$^2$ will cover the LMC, $45$ deg$^2$
the SMC, $20$ deg$^2$ the Bridge and $3$ deg$^2$ the Stream
(Fig.~\ref{vmc}). Observations will be obtained in service mode when
the observing conditions necessary to meet the survey goals are
met. In practice, the most crowded regions like the bars of each
galaxy will be observed when the seeing is $\sim 0.6^{\prime\prime}$
while less populous and outer regions will be observed with a seeing
of $0.8-1.0^{\prime\prime}$.

The sensitivity that the VMC survey will reach is going to match
existing observations at optical wavelengths and will also constitute
their unique near-in\-fra\-red counterpart. These limits, for S/N$=10$,
are: $Y=21.9$, $J=21.4$ and $K_s=20.3$ mag. The expected efficiency of
VMC observations is $\sim 80$\%, but this may change after telescope
first light. To accumulate sufficient integration time to reach the
aforementioned sensitivity the plan is to acquire one epoch at each
filter during a given night and to accumulate two subsequent epochs in
$Y$ and $J$ as well as $11$ epochs in $K_s$ during the same semester
for a given field (a tile).


\subsection{VMC science goals}

The main VMC science goals are the determination of the spatially
resolved star formation history and metallicity evolution across the
Magellanic System, the three-dimensional geometry of the system and
the age dependency (empirical and theoretical) as well as the search
for substructures like new clusters and streams. 

The sensitivity of VMC will allow us to employ different indicators to
meet the major science goals. In particular, the survey will reach
sources $6$ mag fainter compared to the 2MASS and DENIS surveys,
currently broadly used to study the stellar content of the Magellanic
Clouds as a whole. These surveys despite covering both galaxies
homogeneously do reach only upper red giant branch stars. VMC will
include the entire red giant branch population, short period variables
and in particular old RR Lyrae stars, down to the oldest turn-off
stars. Simulations show that this is the required sensitivity to
determine the age of the stellar populations with a resolution of
$0.2$ dex with $20$\% errors.

\begin{figure}[h]
\begin{center}
\includegraphics[scale=0.28, angle=0]{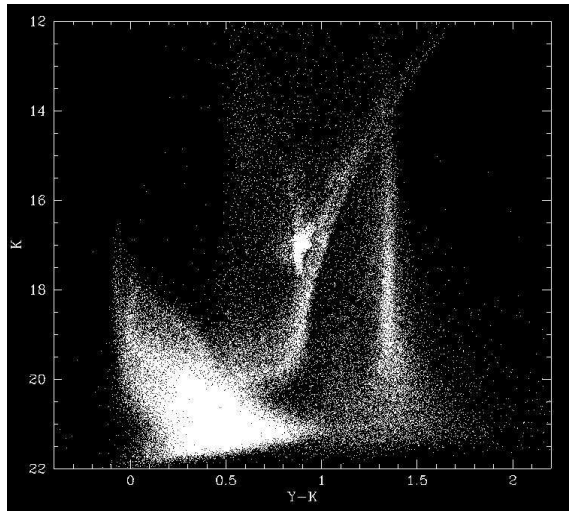}
\caption{Simulated colour-magnitude diagram for a $0.6$ deg$^2$ LMC
  area with a known star formation history. The main sequence, the
  complete RGB, early-AGB and red clump stars are clearly visible; the
  Galactic foreground stars are mostly comprised in an almost vertical
  sequence red-wards of the majority of LMC stars.}
\label{vmccmd}
\end{center}
\end{figure}

The geometry of the system will be measured using the luminosity of
red clump stars, the period-luminosity (PL) relation of RR Lyrae stars
and Cepheids and using standard candles in clusters. All these
indicators are not free from {\it problems} but we expect to produce a
convincing measurement by combining their results.

\subsubsection{Star formation history}

Information about the star formation history (SFH) of the LMC has been
derived from the study of many relatively small regions located in the
outer and inner disk as well as along the bar. In the SMC, apart from
the comprehensive study by \cite{har}, which does not probe the outer
structure, there have been considerably fewer (and less detailed)
observations of the field and cluster stellar populations than in the
LMC. Using VMC data the distribution of field stars in different
phases of evolution will be traced out to distances never yet explored.
In particular we aim to sample the population of RGB stars because
their behaviour is better understood and, because they are likely more
metal poor and, they trace the tidally stripped parts of the galaxies
and the extended halo component. Densities of different stellar
objects are strongly correlated with the SFH.

The most powerful tool for quantitatively measuring the SFH in nearby
galaxies is the analysis of colour-magnitude diagrams via objectives
algorithms that search for the composite model that best fits the
observations. A primary target of the VMC survey is to allow the SFH
measurement in the Magellanic System with unprecedented accuracy and
detail, via this kind of analysis. Figure \ref{vmccmd} shows a typical
simulation for the LMC assuming the VMC targeted depth and an input
SFH constant over the $0.1-12$ Gyr range.

VMC data represent a unique, currently missing, counterpart for optical
sources of similar depth, this will provide us with the ultimate
understanding of the SFH across the system.

\subsubsection{Short-period variables}

RR Lyrae stars trace the old ($t>10$ Gyr) stellar component and follow
a PL relation only in the $K$ band \citep{lon}.  Although it is weakly
affected by evolutionary effects, spreads in stellar mass inside the
instability strip, and uncertainties in the reddening correction, it
does depend on metallicity; \citet{dao} show the application of this
relation to refine the distance to the Reticulum cluster in the
LMC. The theoretical calibration of this relation relies strongly on
the ($V-K$) colour.  RR Lyrae stars in the MCs have $K_s\approx
18.0-19.0$ mag and optical data of comparable sensitivity covering
most of the LMC and the SMC from which to derive the period of the
variation, are or will be available from microlensing surveys while
similar data covering the Bridge will be obtained from STEP
(Sec.~\ref{anc}). Thus, it is of prime importance to measure the mean
$K_s$-band magnitude of RR Lyrae stars with the VMC survey.

Cepheids are young or intermediate-age stars ($100$ Myr) which follow
a much narrower PL relation in the $K_s$ band than the corresponding
optical relations and less affected by systematic uncertainties
related to our knowledge of the reddening and metal content
\citep{c00}; the intrinsic accuracy of the PL-metallicity relation is
$\approx0.05$ mag \citep{dao}. The observed properties of RR Lyrae and
Cepheid stars will be compared with updated theoretical work based on
nonlinear convective models of pulsating stars \citep{m03}.  For
Cepheids, the application of theoretical PL and PL-colour relations to
both near-infrared and optical data will allow us to evaluate self
consistently distances, reddenings and metal abundances
\citep{c99,c01}. Moreover, information on the SFH could be inferred
from the application of theoretical period-age and period-age-colour
relations \citep{m06}.

\subsubsection{Stellar clusters}

Despite a wealth of detailed studies, e.g. \citet{hol,har}, it has not
yet been firmly established whether the field star population has
experienced the same, or a similar, SFH as the star cluster systems,
e.g. \citet{hun,dg06,cha,gie,dg07} in either of the Clouds.  By
combining their integrated photometry with resolved stellar population
studies, the MC cluster system offers a unique chance to independently
check the accuracy of age (and corresponding mass) determinations
based on broad-band spectral energy distributions (BB-SEDs).

\citet{and} and \citet{dg06} developed a method that employs multiple
passband observations to obtain simultaneously cluster ages, mas\-ses,
metallicities and extinction values. Based on a minimum of well-chosen
passbands, absolute ages were derived with a precision of 35\% and
relative ages to an order of magnitude better \citep{dg05}. Moreover,
\citet{and} concluded that to both constrain the cluster ages {\it
and} their metallicities independently using BB-SEDs, one would
require high-quality photometry at {\it both} blue {\it and} red
wavelengths. The requisite data quality at the reddest (near-infrared)
wavelengths is lacking at present, yet its inclusion would allow us to
(i) reach higher accuracy in the cluster age determinations (better
than $\sim 20$\%, which would therefore lead to much firmer
statistical and comparative conclusions regarding the cluster-field
star connection, for instance) and (ii) as a consequence, minimise the
occurrence of artefacts in our cluster age-dating techniques
(sometimes referred to as `chimneys'; see e.g. fig. 7 in
\citet{dg06}. In addition, by including near-infrared passbands we
can constrain any metallicity and extinction variations much more
precisely than by using optical data alone \citep{iva,val}.

Wide-field VMC data will produce a complete census of the cluster
population (both the optically visible and the embedded clusters and
associations) that will allow us to draw statistically robust
conclusions; we will properly compare spatial differences within the
Clouds and possibly -- for the first time -- strongly constrain the
shape of the low-mass cluster mass function, see, e.g.,
\citet{dg07}. This will provide the possibly best constraints on the
evolution of the entire cluster mass function, and hence provide us
with a handle on the clusters' potential for longevity.

\subsubsection{Planetary Nebulae}

The census of Planetary Nebulae (PNe) in the MCs is incomplete and
biased. These objects trace the low- and intermediate-mass stellar
evolution, are important extragalactic distance indicators and
contribute to the replenishment of the interstellar medium with new
elements out of which a new generation of stars may form. Deep
wide-field VMC observations will allow us to uncover the missing
number of PNe contributing to the study of their properties and of the
properties of the host galaxy. A bi-product of this research will be
the compilation of the Magellanic Extended Source Selection (MESS)
catalogue.

PNe are recognized as emission line objects in particular of H$\alpha$
and [OIII] but also of other elements. Bright central stars and
nebulae in uncrowded regions of the MCs were imaged by the Hubble
Space Telescope down to $V\sim 25$ and by Spitzer in the
mid-infrared. In the near-infrared VMC will reach a comparable
sensitivity. PN will be bright in $K_s$ because of Br$\gamma$ emission
and much fainter, if detected, in $Y$ and $J$ (continuum) compared to
other emission line objects; Br$\gamma$ is less sensitive to
reddening. The combination with deep optical imaging and spectroscopy
will not only contribute to the selection and identification of new
PNe but will resolve the ambiguity with HII regions, young stellar
objects, SN remnants and background galaxies.

Recent observations have shown that the surface brightness of PNe is
well correlated with size, the fainter PNe tend to be larger and
current samples are highly incomplete at this level. Just the central
$25$ deg$^2$ of the LMC have been surveyed by deep H$\alpha$
observations and that has already tripled the number of previously
known PNe. Observations of the SMC cover a larger area of the galaxy
but are on average shallower. In addition, older objects are
considerably undersampled, these are usually found out to large
distances from the centre. The present data show a spatially and
evolutionary biased sample of PNe which limits a complete
understanding of this late evolutionary phase and a broad use of these
objects (i.e.~luminosity function, progenitors). Considering that the
AGB population of the LMC and SMC define smooth elliptical structures
covering areas of about $116$ deg$^2$ and $45$ deg$^2$, respectively,
there is plenty of room for new discoveries!

\subsubsection{Ancillary goals}
\label{anc}

There are other science topics that VMC will considerably contribute
to. Those in which the team is directly involved are: the
determination of the distance to the LMC (a reference for the distance
scale in the Universe) with an unprecedented quality that will reduce
by a factor of two current uncertainties; finding obscured massive
stars and unreddened $1.5$ M$_{\odot}$ pre-main sequence stars;
determine the proper motion of the Magellanic Clouds by using VMC data
alone over the five years of the survey, resulting in an accuracy of
$\sim 0.05$ mas/yr, or by combining VMC data with 2MASS data spanning
a total time of about $15$ years, reducing further the uncertainty in
the measures. Finally, one of the most important aspects of VMC will
be to provide targets for new follow-up observations using telescopes
with a larger collecting areas like the Very Large Telescope (VLT).

\subsection{VMC complementary surveys}

The VMC data alone will be a valuable resource which aims to explain
the Magellanic System but also to provide a highly valuable
counterpart for stars detected by other means, building on the
scientific value of the research each survey set out to produce. Among
these complementary surveys are:

\begin{itemize}
\item EROS, OGLE and MACHO microlensing sur\-veys that observed and/or
 are still observing extended areas centred on each MC periodically
 for/since several years allowing the characterisation and discovery
 of many short period variables such as RR Lyrae and Cepheids, these
 data will be used to measure the period of these variables while
 their mean $K_s$ magnitude will be obtained by VMC, these
 period-magnitude relations will be used to trace the
 three-dimensional geometry of the MCs;
\item SIRIUS and deep-2MASS are two near-infrared surveys of the MCs
which will reach a magnitude fainter sources compared to the original
2MASS survey, these surveys have been completed very recently and will
represent a step further in the study of near-infrared stellar
populations in the MCs until the much deeper VISTA data will become
available;
\item SAGE and S3MC are two surveys currently on-going from the
Spitzer Space Telescope, they focus on obscured sources like embedded
late-type stars and star forming regions in the MCs, many of these
objects do not have a near-infrared counterpart and are too obscured
to be detected at optical wavelength, this aspect severely limits
their classification and study and VMC data will therefore be
extremely valuable;
\item MOSAIC is a deep optical survey of the outer MCs from
$7^{\circ}$ to $20^{\circ}$ from the centre, the overlap with VMC data
will be minimum but both surveys will be highly complementary;
\item AKARI is a mid-infrared space telescope which is currently
  observing the MCs via an all sky survey but also via dedicated
  observations. These data, analogous to the Spitzer data, will 
  characterise dusty sources for which VMC counterparts will prove
  highly useful.
\item STEP is an optical survey of the SMC and the Bridge components
  of the Magellanic System to be performed with the VLT survey
  telescope (the VST), these data will provide unique periods for short
  period variables located in the Bridge and will also reach sources
  as faint as VMC will do across the SMC;
\item GAIA is an astrometric survey which is due to begin within the
  next decade, it will measure the accurate proper motion of the MCs,
  the kinematics and metallicity (via the Ca II triplet) of all bright
  giant stars, these informations combined with VMC data as well as
  with optical data of a similar sensitivity will provide the
  necessary ingredients to study the evolution of the Magellanic
  System, the propagation of star formation as well as the interaction
  with the Milky Way galaxy.
\end{itemize}  

\section{Conclusions}

The Magellanic System has yet many challenging aspects that new
surveys, with the increased quality of the coming data and 
new theoretical models and their ability to explain detail
observations, aim to resolve in the next decade. 

Prior to new facilities like GAIA, JWST and ALMA we need to exploit
data from VISTA and similarly powerful telescopes at other
wavelengths. Surveys like VMC will provide unique and high quality
data for science and training of young astronomers.

\section*{Acknowledgments} 

Cioni would like to thank the organizers of the Elizabeth and
Frederick White Conference on the Magellanic System for inviting her
to present this key talk.


\end{document}